\documentclass{jps-cp}
\usepackage{txfonts} 

\usepackage{graphicx}
\usepackage{comment}
\usepackage{floatrow}

\title{Propagation of QCD Color through Strongly
Interacting Systems}

\author{William K.  \textsc{Brooks}$^{1}$ and Jorge A.  \textsc{L\'opez}$^{1,2}$}

\inst{$^{1}$Universidad T\'ecnica Federico Santa Mar\'ia, Avenida Espa\~na 1680 Casilla 110-V, Valpara\'iso, Chile \\
$^{2}$Physikalisches Institut, Ruprecht-Karls-Universit\"at Heidelberg, Heidelberg, Germany}

\email{william.brooks@usm.cl}

\recdate{February 1, 2019}

\abst{The propagation of QCD color through atomic nuclei is studied via a new analysis using a geometric model of semi-inclusive deep inelastic scattering. The experimental data were previously published by the HERMES Collaboration and consisted of the multiplicity ratio observable (2007) and the transverse momentum broadening observable (2010). We perform a simultaneous fit of these two observables to estimate (1) the color lifetime of the quark, (2) quark energy loss in the nuclear medium, (3) the ${\hat{q}}$ transport coefficient, and (4) the cross section for hadronic interaction with the medium. We present preliminary results for this fit.}

\kword{parton energy loss, nuclear DIS, space-time QCD}

\begin{document}
\maketitle

\section{Introduction}
Experimental studies of the propagation of color charge through strongly interacting systems probe fundamental features of QCD in new ways. The physics associated with these studies constitutes the foundation of color charge propagation and hadron formation in vacuum, and of jet quenching in heavy ion physics; and the fundamental processes involved are of high interest for cold matter studies using semi-inclusive deep inelastic scattering (SIDIS) kinematics at Jefferson Lab and at the future Electron-Ion Collider. Despite two decades of experimental and theoretical effort, there are very few areas in which precise and quantitative confirmation of the theoretical ideas has been unambiguously observed, such as in the study of quark energy loss in atomic nuclei. In this work, we use a geometric model of nuclear densities and parton trajectories to simultaneously fit two observables in the HERMES SIDIS data on neon, krypton, and xenon nuclei compared to light nuclei. With this fit we are able to estimate the energy loss and color lifetime of the quark, the ${\hat{q}}$ transport coefficient, and the effective cross section for hadronic interaction with the medium.

\section{Definition of the Observables and of the Model}
Transverse momentum broadening $\Delta p_\mathrm{T}^2$ is the increase in the mean value of the $p_\mathrm{T}$ spectrum in a larger nucleus A compared to a smaller nucleus D: 
\begin{equation}
\Delta p_\mathrm{T}^2 (Q^2,\nu,z_\mathrm{h}) \equiv \left\langle p_\mathrm{T}^2(Q^2,\nu,z_\mathrm{h}) \right\rangle \rvert_\text{A} -  \left\langle p_\mathrm{T}^2(Q^2,\nu,z_\mathrm{h}) \right\rangle \rvert_\text{D}
\label{eq:ptb_d}
\end{equation}

The multiplicity ratio $R_\mathrm{M}^\mathrm{h}$ measures the attenuation or enhancement of hadrons in a larger nucleus "A" compared to a smaller nucleus "D":
\begin{equation}
R_\mathrm{M}^\mathrm{h} (Q^2,\nu,z_\mathrm{h},p_\mathrm{T}) \equiv \frac{\dfrac{N_\mathrm{h} (Q^2,\nu,z_\mathrm{h},p_\mathrm{T})}{N_e(Q^2,\nu)} \bigg\rvert_\text{A}}
                                    {\dfrac{N_\mathrm{h} (Q^2,\nu,z_\mathrm{h},p_\mathrm{T})}{N_e(Q^2,\nu)} \bigg\rvert_\text{D}}
\label{eq:mr_d}
\end{equation}
where $N_{e(\mathrm{h})}$ is the number of electrons (hadrons) observed in the kinematic bin indicated. This observable is equal to unity in the absence of all nuclear effects. In the pion data used in this study, $R_\mathrm{M}^\mathrm{h}$ is generally less than unity, i.e., a suppression of hadrons is observed.

We interpret the $p_\mathrm{T}$ broadening as a consequence of a color-charged object undergoing multiple scattering as it passes through a medium: 
\begin{equation}
\langle \Delta p_\mathrm{T}^2 \rangle = \left\langle q_0 \int_{z_0}^{z_0+L_\mathrm{c}^*}\rho (x_0,y_0,\tilde{z})\text{d}\tilde{z} \right\rangle _{x_0,y_0,z_0,L_\mathrm{c}}
\label{eq:ptb_m}
\end{equation}
where $(x_0,~y_0,~z_0)$ is the coordinate of the hard interaction point, $q_0$ is a fit parameter related to the $\hat{q}$ transport coefficient, $L_\mathrm{c}$ is a fit parameter representing the characteristic color length, $L_\mathrm{c}^*$ is the lesser of $L_\mathrm{c}$ or the distance from $z_0$ to the sphere of integration surface, and $\rho$ is the nuclear density function taken from Ref.\cite{henk}. The $\tilde{z}$ variable is the longitudinal coordinate over which the density is integrated.

In the baseline model BL we interpret the attenuation of hadrons represented by the multiplicity ratio as being caused by a hadronic interaction of the forming system (pre-hadron) that contains the struck quark, and thus we associate it with an effective hadronic cross section: 
\begin{equation}
\langle R_\mathrm{M}^\mathrm{h} \rangle = \left\langle \mathrm{exp}\left(-\sigma\int_{z_0+L_\mathrm{c}^*}^{z_\text{max}}\rho (x_0,y_0,\tilde{z})\text{d}\tilde{z} \right) \right\rangle _{x_0,y_0,z_0,L_\mathrm{c}}
\label{eq:mr_m}
\end{equation}
where the symbols are as defined for the previous equation, $\sigma$ is the effective prehadron-nucleon cross section, and $z_\text{max}$ is the maximum value of the coordinate $\tilde{z}$ that is still inside the integration sphere. The model is implemented as a Monte Carlo calculation.

Another mechanism which can modify the multiplicity ratio is quark energy loss, which results in a migration of produced hadrons in larger nuclei toward smaller energies and thus smaller $z$, where $z \equiv {p_\mathrm{h}^+/{p^+}}$ is the light cone momentum of the hadron divided by the total light cone momentum. The variable $z$ has a similar value to the variable $z_\mathrm{h} \equiv {E_\mathrm{h}/{\nu}} $ which was used in the HERMES publications, but has a better representation on the light cone (see Appendix). We implement quark energy loss as a fitted shift in $z$ toward lower values in a variant of the baseline model shown in equations \ref{eq:ptb_m} and \ref{eq:mr_m}, labeled as BLE30.

\section{Data Treatment}
The HERMES data that we fitted to our model were taken from the positive pion data in the final one-dimensional multiplicity ratio paper \cite{HERMES2007} and the transverse momentum broadening paper \cite{HERMES2010}. The data binning of the latter paper was used for the fit, and the data in the former paper were interpolated to provide values in the same $z_\mathrm{h}$ bins.

\section{Results}
\begin{figure}[tbh]
\includegraphics[width=0.49\linewidth]{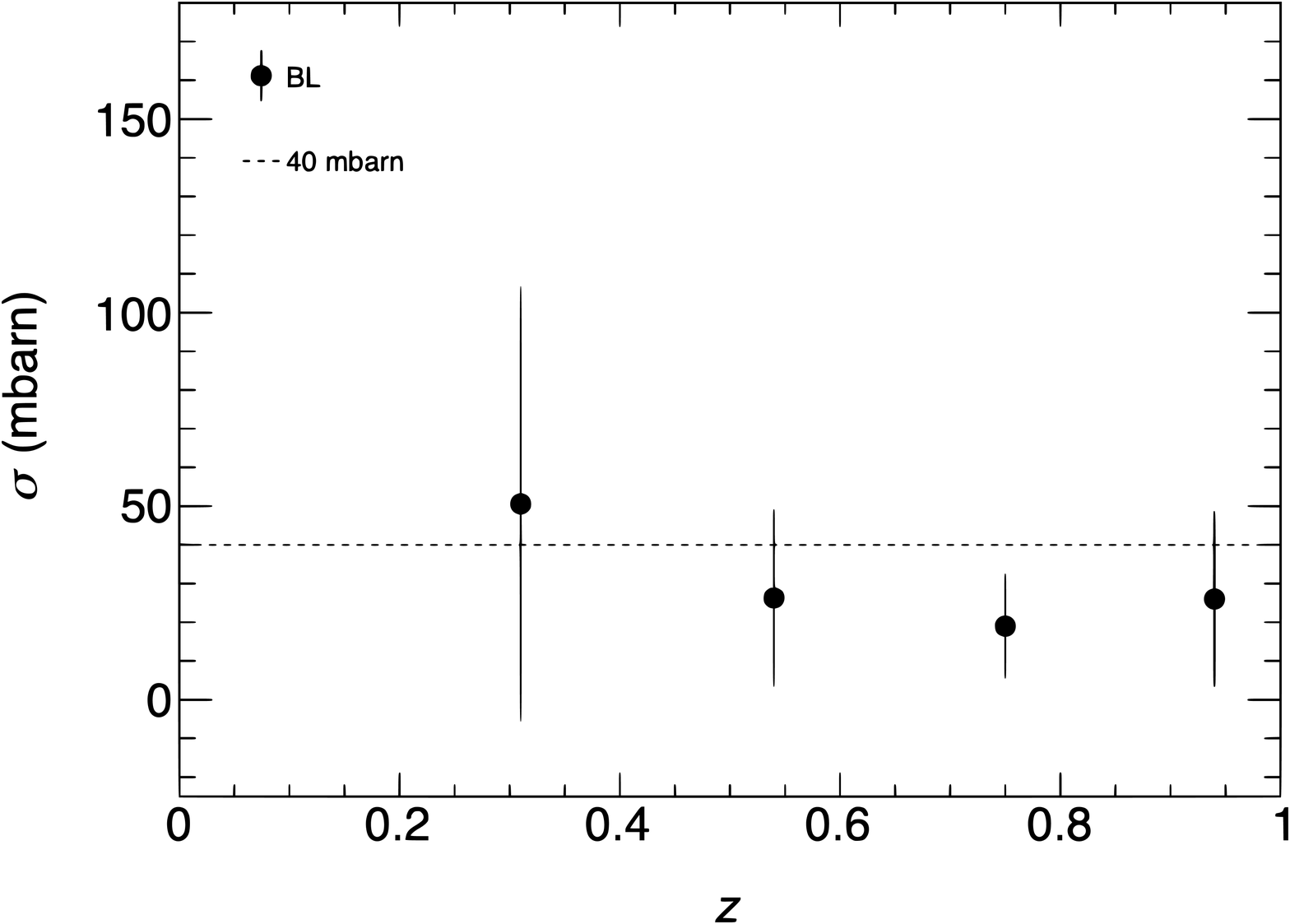}
\includegraphics[width=0.49\linewidth]{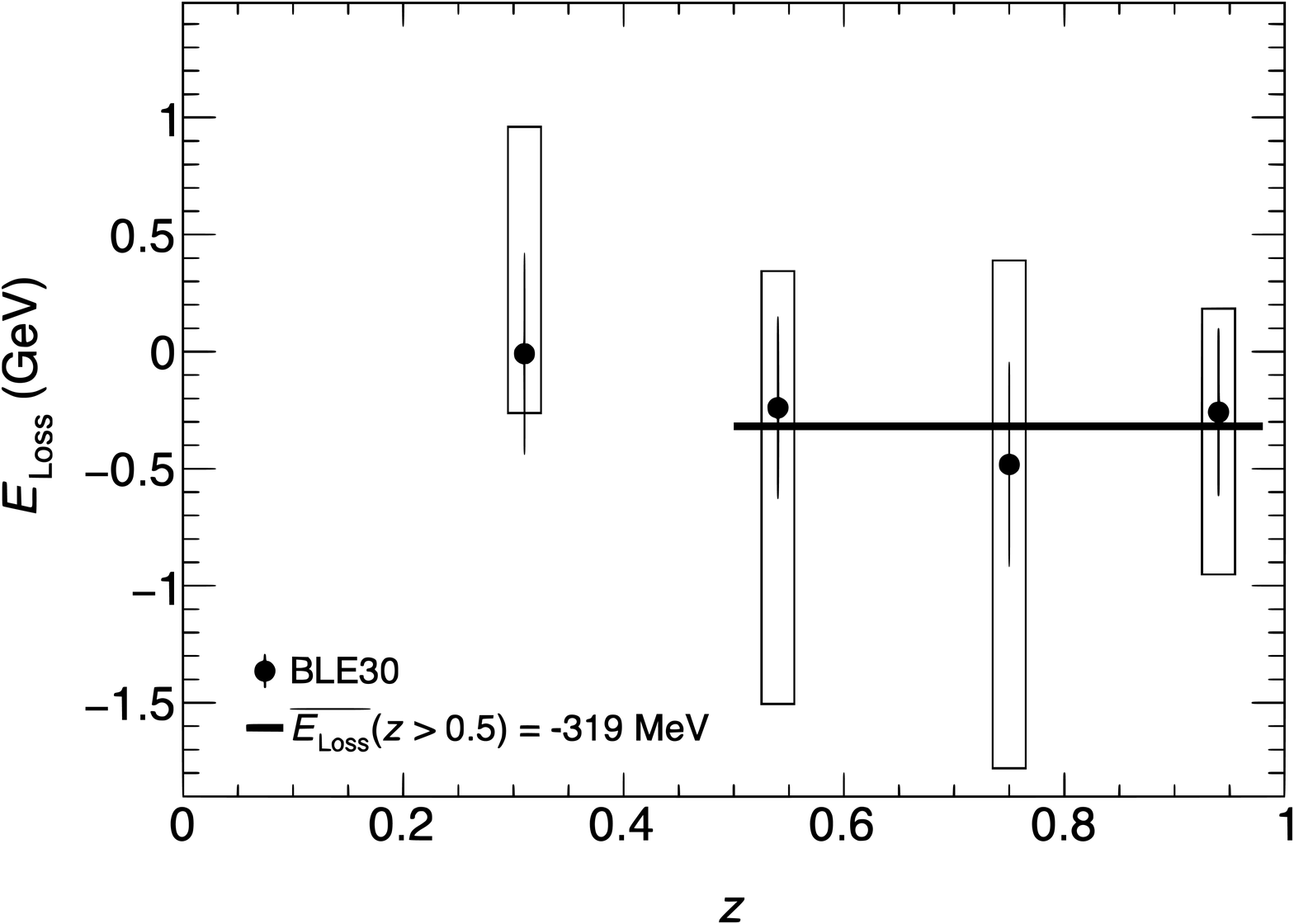}
\caption{Results of the simultaneous fit of the model to two observables from the HERMES publications. The results are presented as a function of  $z$, the fraction of light cone momentum carried by the hadron. (Left) Effective hadronic cross section from the baseline model given in equations \ref{eq:ptb_m} and \ref{eq:mr_m}. The error bars show the fit uncertainties. A reference line from pion-nucleon scattering is shown at 40 mb. Since the pions in this measurement are in formation, it is anticipated that the cross section should be smaller than that of fully formed pions, consistent with the fit results for $z>0.5$ in the figure. (Right) Quark energy loss derived from a variant of the baseline model in which energy loss can contribute to the suppression of pions as measured by the multiplicity ratio, and for which the effective hadronic cross section is fixed at 30 mb. The average value for $z>0.5$ is shown. The error bars show the fit uncertainties, while the boxes show the range of values found from the baseline model and five model variants. }
\label{fig:results}
\end{figure}

In our model we avoid dynamical assumptions, and instead perform fits in each of four bins in $z$, constrained by the known geometrical density distributions of the nuclei. By doing this we extract the dynamical behavior of the four quantities of interest as a function of $z$. For example, for the HERMES data we find a color lifetime that decreases from approximately 8 fm/c to approximately 2 fm/c as $z$ ranges from 0.3 to 0.9 with a functional form very similar to that of the Lund String model (see Appendix). We observe essentially no $z$ dependence of the quark energy loss shown in Fig.\ref{fig:results} (right), nor in the hadronic interaction cross section shown in Fig. \ref{fig:results} (left), and similarly the $\hat{q}$ values we find are nearly independent of $z$.

\section{Conclusions}
A geometric model has been used to analyze the SIDIS production of positive pions in the HERMES data. A simultaneous fit to two observables has permitted preliminary estimation of  (1) the color lifetime of the quark, (2) quark energy loss, (3) the ${\hat{q}}$ transport coefficient, and (4) the cross section for hadronic interaction with the medium. A $z$ dependence of the color lifetime is observed, while the other quantities are consistent with being constant in $z$. The cross section results are consistent with being smaller than the typical hadron-nucleon cross section, as would be anticipated for a forming hadron.

\section{Acknowledgements}
We acknowledge valuable discussions with T. Sj\"ostrand, B. Kopeliovich, S. Brodsky, A. Majumder, and S. Peigne. This work was supported by Chilean grants PIA ACT-1413, ACT-119, and BASAL FB-0821; Fondecyt 1080564, 1120953, and 1161642; PAI ECOS-CONICYT C12E04; Beca Doctorado Nacional 2014 21140777; and UTFSM Beca DGIIP-PIIC. This work was initiated from discussions in 2009 during the Institute for Nuclear Theory program INT-09-3 and workshop INT-09-43W.

\appendix
\section{Derivation of the Lund String Model Form of the Color Lifetime}
\begin{figure}[tbh]
\floatbox[{\capbeside\thisfloatsetup{capbesideposition={left},capbesidewidth=8cm}}]{figure}[\FBwidth]
{\caption{Space-time diagram to illustrate the discussion of this Appendix. The time axis is vertical in the diagram, while the displacement $\tilde{z} $ axis is in the horizontal direction. The first vertex is indicated, and the energy-momentum fraction $z$ of the forming hadron is indicated as a fraction of the total energy-momentum $P^+$ available from the initial interaction. The subsequent formation of other hadrons is not shown in the figure. Thus, Eq. \ref{eq:eq5} essentially corresponds to the struck quark, in pQCD terminology.}\label{fig:Lund}}
{\includegraphics[width=6cm]{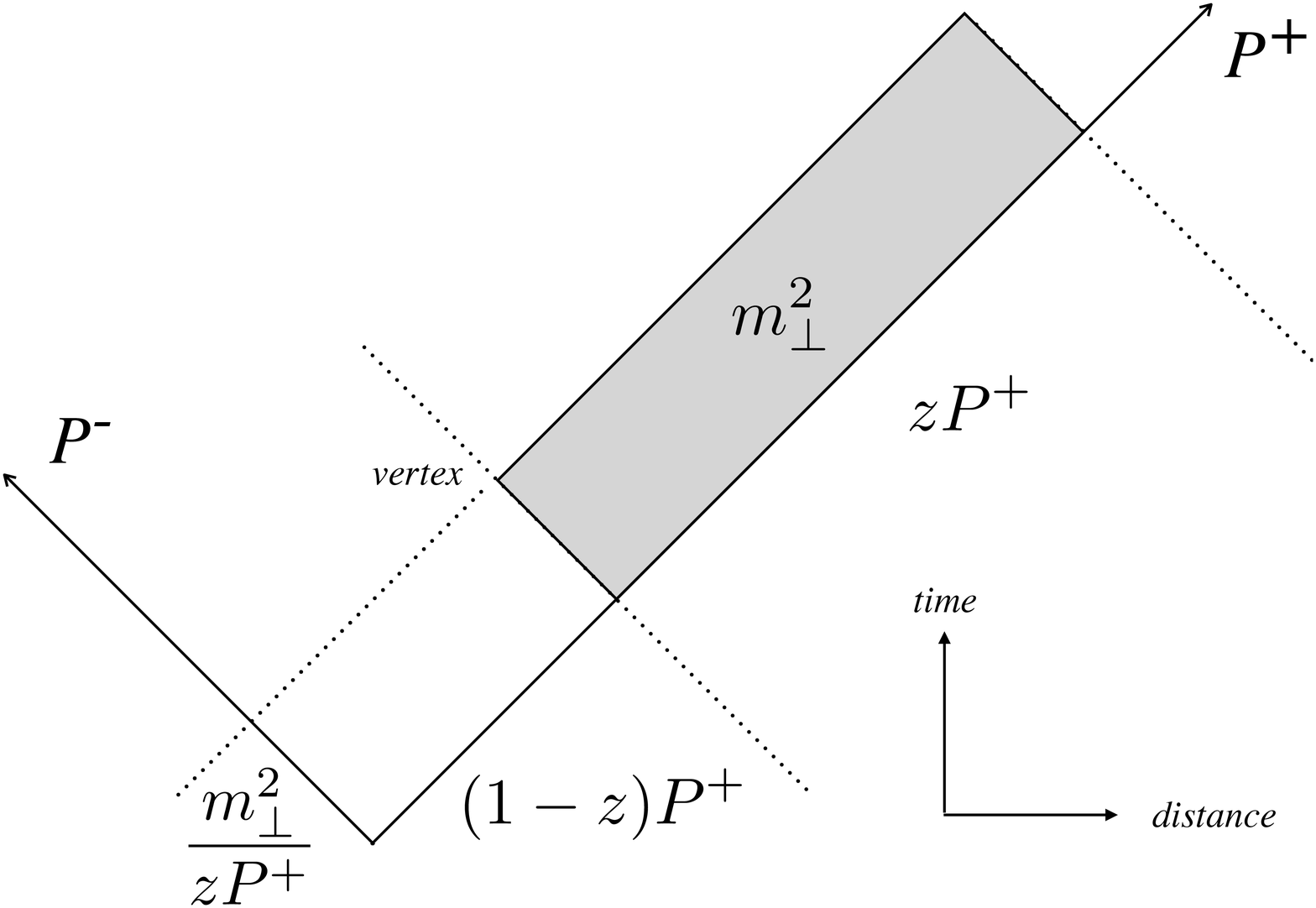}}
\end{figure}

\noindent Consider a proton of mass $M$ initially at rest, and an incoming photon in the $\tilde{z}$ direction with energy $\nu$ and momentum $p_z = \nu\sqrt{1+Q^2/\nu^2}\approx\nu$, given that $Q^2/\nu^2\ll1$, where $Q^2$ is the four-momentum transferred by the photon. Expressing the dynamics on the light cone, the components $P^\pm=E\pm p_z$ of the total four-momentum vector can be written:
\begin{eqnarray}
\label{eq1}
P^+ =& P^+_\mathrm{proton} + P^+_\mathrm{photon} =& M + 2\nu \\
P^- =& P^-_\mathrm{proton} + P^-_\mathrm{photon} =& M
\end{eqnarray}
A hadron being formed over some distance from the scattering point satisfies the condition that $p^+_{h}p^-_{h}=m_\perp^2$ \cite{Andersson:1983ia}. The forming hadron carries a fraction $z$ of the available positive momentum $P^+$, such that $p^+_{h}=z P^+$ (see Fig. \ref{fig:Lund}), and thus $p^-_{h} = m_\perp^2/(z P^+)$. The remaining available momentum moves with the vertex, which has propagated a distance $L$ from the interaction point over a time $T$. We therefore have the following equations:
\begin{eqnarray}
p^+_{\text{vertex}} =& \kappa (T+L) =& (1-z) P^+ \\
p^-_{\text{vertex}} =& \kappa (T-L) =& p^-_\mathrm{h} 
\end{eqnarray}
\noindent where $\kappa$ is the Lund string constant. Solving for $L$ and using Eq. \ref{eq1} for $P^+$:
\begin{equation}
\label{eq:eq5}
L = \frac{1}{2\kappa}\left((1-z) (M + 2\nu) - \frac{m_\perp^2}{z (M + 2\nu)}\right)
\end{equation}

It is also possible to connect the fraction of light cone momenta carried by the hadron $z$ with the usual definition of this fraction in DIS: $z_\mathrm{h}= E/\nu$. The experimental energy fraction is $z_\mathrm{h}=E_\text{hadron}/\nu$, while the light cone fraction is $z=p_{h}^+/p^{+}$:
\begin{equation}
z = \frac{p_{h}^+}{p^{+}} = \frac{E_\mathrm{h} + p_{\mathrm{h},z}}{M + 2 \nu} = z_\mathrm{h} \frac{1+\sqrt{1-m_\perp	^2/(z_\mathrm{h}\nu)^2}}{M/\nu + 1 + \sqrt{1 + Q^2/\nu^2}}
\end{equation}

\end{document}